\documentclass{iosart2c}

\usepackage[T1]{fontenc}
\usepackage{times}%

\usepackage{amsmath}
\usepackage{dcolumn}
\usepackage{graphics}
\usepackage{graphicx}

\newcolumntype{d}[1]{D{.}{.}{#1}}

\firstpage{1} \lastpage{10} \volume{1} \journal{VU University}  \pubyear{2016}

\begin{document}
\begin{frontmatter}

%
\title{The Entity Registry System. From concept to deployment}


\author[A]{\fnms{Mihai} \snm{Gramada}\thanks{Corresponding author. E-mail: gramada.mihai@gmail.com.}},
\runningauthor{Mihai Gramada}
\runningtitle{ERS. From concept to deployment.}
\address[A]{VU University Amsterdam \\
E-mail: gramada.mihai@gmail.com}
\author[A]{Supervisor: \fnms{Christophe} \snm{Gueret} \thanks{Supervisor. E-mail: christophe.gueret@bbc.co.uk} }
\author[A]{Second Reader: \fnms{Victor} \snm{De Boer} \thanks{Second reader. E-mail: v.de.boer@vu.nl}}

\begin{abstract}
The entity registry system (ERS) is a decentralized entity registry that can be used to replace the Web as a platform for publishing linked data when the latter is not available. In developing countries, where off-line is the default mode of operation, centralized linked data solutions fail to address the needs of the communities. Although the features are mostly completed, the system is not yet ready for deployment. This project aims to provide extensive tests and scalability investigations that would make it ready for a real scenario.
\end{abstract}

\begin{keyword}
Linked data\sep Decentralized System \sep ICT4D \sep RDF
\end{keyword}

\end{frontmatter}

\section{Introduction}

Applications tend to consume and produce more and more data with each passing year. In order to cope with this increasing volume, some of the work of interpreting and connecting the various pieces can be shifted to the machines that process it. Linked data aims to accomplish this by providing some structure to the data. This can be especially useful for social networks, wikis or learning systems, where the information is highly interconnected. Linked data stores are already numerous and massive in size. For example the DBpedia data set is estimated to contain 4.5 million entities \footnote{\text{http://wiki.dbpedia.org/about}}
\begin{figure}[ht!]
\centering
\includegraphics[width=60mm]{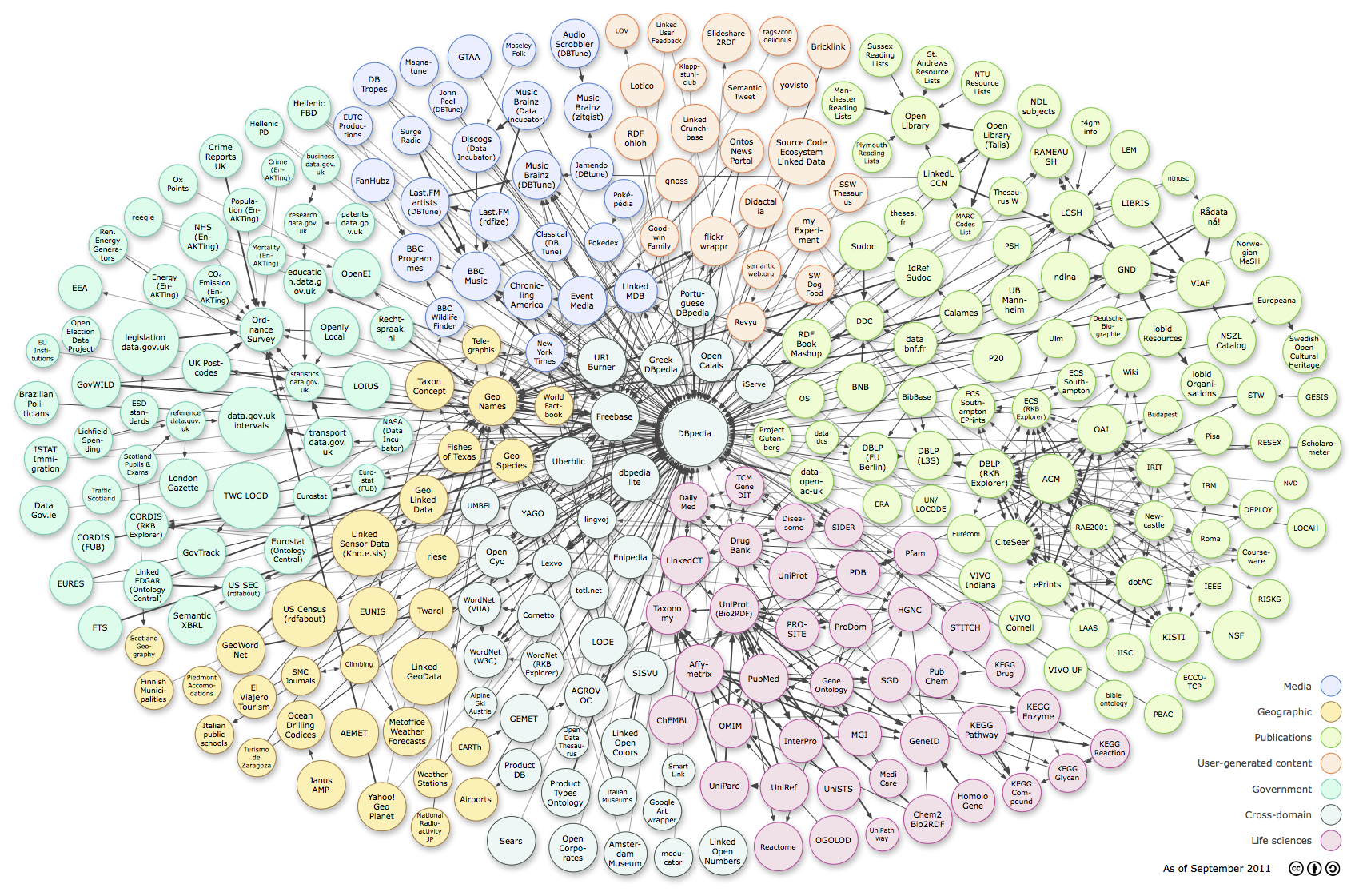}
\caption{Linked open data cloud diagram \cite{loddiagram} \label{overflow}}
\end{figure}
Most linked data applications utilize a central repository that is always online and can be interacted with by the users. However, always-online connectivity is not guaranteed in developing countries, especially outside of the major cities. Current estimates indicate that approximately 4.1 billion people live with limited to no internet access\footnote{\text{ http://www.internetlivestats.com/internet-users-by-country/2015/}}. Information and Communication Technologies for Development (ICT4D \footnote{\text{ http://www.ict4dc.org/about}}) tries to improve the quality of life of those in developing countries by providing access to technology \cite{ict4d}.

In order to address the issue of deploying applications that utilize linked data in limited connectivity circumstances, under a grant from Verisign\cite{ersdevs}, the Entity Registry System (ERS) has been developed. Its goal is to replace the Web as a platform for publishing Linked Data when the latter is not available. It proposes a completely decentralized model, where off-line is the default mode of operation, with occasional connectivity bursts.

An entity registry provides a way to associate data with an uniquely identifiable entity. ERS is a decentralized, read-write entity registry that allows collaborative editing of entities, which have their data represented in the Resource Description Framework (RDF \footnote{\text{https://www.w3.org/TR/2004/REC-rdf-concepts-20040210/}} \cite{RDF} format. By using RDF tuples as the underlying data representation, ERS can inter-operate with existing linked data sets, providing a robust system that can be used both online and offline, can be deployed on very low-end hardware as well as high-performance servers and has a high degree of tolerance for sudden network and topology changes.

The goal of this work is to investigate ERS behavior in complex realistic scenarios. To this end, we needed to make it easy to deploy, configure and orchestrate, in order to create repeatable test scenarios. We also verify that the system functions without ideal network connectivity and low hardware capabilities.

Section two describes the work related to ERS, and also the tools that are used when automating systems testing. Section three describes ERS as it was at the beginning of the work. The following section describes the components of the testing framework and the improvements and changes that needed to be made to the original system. Section five describes the various experiments that were run with the developed tools and is followed by a discussion section that describes the lessons learned. Finally, the conclusion section also investigates possible future improvements.

\section{Related work}

\subsection{Data storage systems}
In \cite{christophemain} an overview of the options for designing systems that can store and manipulate linked data is given, and the three main categories are: centralized, hierarchical and distributed. A centralized system depends on a central node or group of nodes for storing and serving data, causing the central component to become a single point of failure for the entire system. A hierarchical architecture mitigates this problem by providing some functionality from nodes outside of the central cluster, whereas a fully distributed approach distributes the load across all constituent nodes.

\begin{itemize}
\item centralized :
 \textbf{OKKAM\cite{okkam}} and \textbf{ConceptWiki\footnote{\text{http://www.conceptwiki.org/}}} are two web applications that allow the manipulation of entities. However, due to their centralized model, they are not designed to function in a poor connectivity scenario.

\item hierarchical :
\textbf{DNS} is the canonical example of a registry which uses the hierarchical architectural model. In \cite{christophemain}  the possibility of extending the DNS system by using the text record field and DNSSEC to include meta-data about internet domains is suggested.

\item decentralized :
In \cite{ersdevs} an overview of several decentralized platforms that are similar to ERS is given.

	\textbf{ TIPC (Transparent Inter-process Communication)} \footnote{\text{http://tipc.sourceforge.net/}} \cite{tipc} is a network protocol with many architectural similarities to ERS. The architecture consists of nodes, clusters and zones. Nodes can establish point-to-point links to each other (in a full mesh) that are logically grouped into clusters, which in turn can be grouped into zones. There are some differences to ERS. First of all, in TIPC the grouping is logical, whereas ERS provides different components for the nodes, bridges(which correspond to clusters) and zones (which correspond to the global aggregator). Secondly, the behavior is different. If ERS nodes within a network detect a bridge, they switch connectivity from full mesh to connecting only to the bridge, thus reducing the number of links from quadratic to linear. Finally, TIPC functions at the IP layer, whereas ERS functions at the application layer of the network stack.

	\textbf{ Sugar Network} \footnote{\text{https://wiki.sugarlabs.org/go/Platform\_Team/Sugar\_Network/Architecture}} is a project that is very similar to ERS. However, ERS proposes a decentralized model that will work in completely closed networks, with nodes sharing information without the presence of bridges, whereas in the Sugar Network architecture, clients can only communicate with nodes (which are the equivalent of bridges) or the Master node (which are the equivalent of the global aggregator). The data model of ERS also permits more flexibility, as it allows storing of any data as long as it is expressed as RDF tuples.

    \textbf{ Nintendo StreetPass and SpotPass} \footnote{\text{https://en.wikipedia.org/wiki/SpotPass\_and\_StreetPass}} represent systems designed to allow users of the 3DS handheld game system to share game data seamlessly, when in close proximity. This functionality is similar to that of ERS nodes being connected on the same network. There is also the option of using StreetPass relays which are similar in functionality to ERS bridges. The main difference is that the ERS design also features a top-level aggregator that can offer snapshots of the entire system. Moreover, ERS is not restricted to a particular platform (deployments have been tested on XO laptops, raspberry pis and similar boards, and full x86 systems).

\end{itemize}

The centralized and hierarchical models provide a poor fit for the scenario which ERS targets, as their design architecture is not meant for ad hoc mesh network connectivity. The existing decentralized solutions provide either strong restrictions on the devices that are used (e.g. Nintendo StreetPass) or the data model (e.g. Sugar Network). Though TIPC provides a similar architecture, it is meant to function as part of the network stack. ERS functions at the application layer and provides clear deployment scenarios and is designed for ad hoc mesh networks in remote areas.

\subsection{Testing infrastructure}

With the increasing popularity of large scale distributed systems and cloud computing,  automating the management of infrastructure (Infrastructure as Code) has become a recommended practice\cite{infrastructure as code}. The devops role has greatly benefited from the appearance of tools that assist with provisioning, hardware management, deployment, and failure simulation.

Provisioning tools such as Puppet\footnote{\text{https://puppet.com/product/how-puppet-works}} or Chef\footnote{\text{https://www.chef.io/chef/}} help reduce the risk of human errors and can thus offer great end to end time improvements when used in conjunction with cloud providers that can offer physical resources within minutes. When a system requires new hardware, it can be automatically configured and added to the pool of available resources. These tools can help reduce the ERS setup time drastically when running natively.

Docker\footnote{\text{https://www.docker.com/what-docker}} allows developers to package their applications into containers that can be run on any supported hardware. Having pre-built docker containers for ERS available online makes it easier to test . A new instance of ERS can thus be set up in minutes.

One of the most innovative additions to the field was the Netflix simian army \footnote{\text{http://techblog.netflix.com/2011/07/netflix-simian-army.html}}. In the wake of a major outage, Netflix has developed a suite of tools that allows them to simulate multiple types of failure without human intervention. Having this as an automatic process helps prevent testing only the cases that the developers have thought of. Also, having failure as a constant process means that automated recovery becomes a critical part of the system, and when a real failure happens, the users do not notice it.
Although Netflix has released the tools as open source software, it is designed to function in conjuncture with AWS and is not usable locally, and the configuration effort is also pretty significant.

Section four describes how ERS uses Puppet for provisioning, Docker containers, as well as an implementation of the Netflix Simian Army that can be run without AWS. Section five describes how these building blocks were used to create repeatable, complex test scenarios, that can run without manual intervention for the purpose of investigating ERS functionalities, performance and scalability.

\section{The Entity Registry System}

\subsection{Overview}
\begin{figure}[ht!]
\centering
\includegraphics[width=85mm]{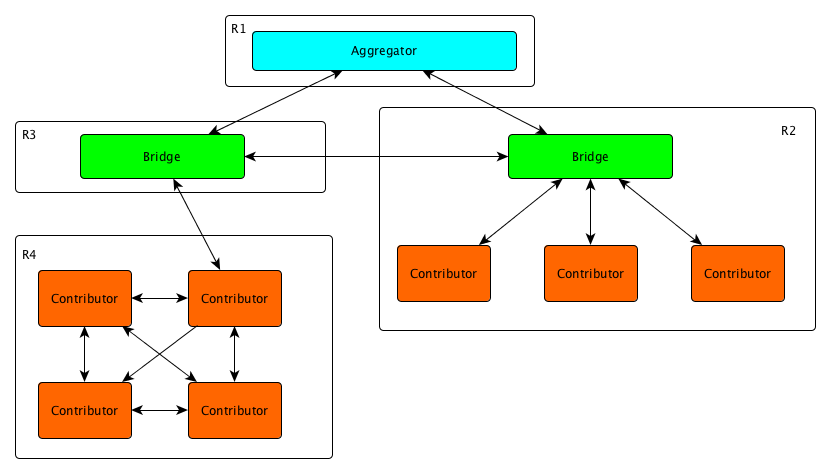}
\caption{Sample deployment with 4 geographically separate regions(R1-R4). All nodes can continue functioning while connectivity is disrupted, and updates can propagate when connectivity is restored between the regions.\cite{ersdevs}\label{overflow}}
\label{fig:deployment}
\end{figure}

In Figure\ref{fig:deployment} a simple example of an ERS deployment can be observed. More details about the system can be seen on the Github page\cite{ersdevs} For more in-depth analysis, please refer to the paper describing the system \cite{christophe1}. There are three main components:

\begin{itemize}

\item Contributor: The contributors can read and write the content of the registry. In the general use case, they would be represented by the laptops of the children, with the content of the registry being the linked data that various activities produce. Examples of such activities may be contributions on a shared document, notes on various book chapters, etc. This data is stored locally and can be distributed (if made public) whenever connectivity is present.

\item Bridges: Bridges provide a means to connect different isolated parts of the system in order to distribute public data from the contributors. They do not need to reside in the same physical location as any of the contributors and can share public data provided to them with bridges in any other locations. Their functionality is similar to that of a cache, in the sense that if a part of the network goes down, public data can still be read from the bridges to which it was previously connected. Of course, due to storage constraints, data is stored only temporarily on the bridge.

\item Aggregator: ERS deployments can benefit from the presence of an aggregator, which would ideally be deployed on a high-performance server or cluster. This optional component can provide a top-level entry point to the whole system, enabling fast read-only retrieval of the public data. Contributors and bridges can push data to the aggregator for it to be available publicly.

The example in Figure\ref{fig:deployment} shows a deployment with contributors and bridges in multiple geographical regions (R1 through R4). In this example, R1 contains the aggregator, which can offer a read-only view of the public contributions of all the contributors. In R4 we see a number of contributors without a bridge present on the network, and thus each contributor synchronizes with all the others. One of the contributors in R4 is set to send updates to the bridge in R3 whenever it is available. In contrast, R2 has a bridge present on the network and all the contributors share data through it instead of directly to each other. The bridges are set to synchronize with each other and with the aggregator.
In this example, when full connectivity is present, public data can be read by any contributor, or can be queried from the aggregator. Should R1 go offline, the rest of the nodes will not be affected. The system is resilient towards arbitrary failures, with nodes and bridges being able to connect and disconnect without heavily impacting others. The nodes will only have stale data while connectivity is fully disrupted between regions. Once it is re-established, the synchronization protocol will handle propagating the updates.
\end{itemize}

\subsection {Data representation}
\begin{figure}[ht!]
\centering
\includegraphics[width=60mm]{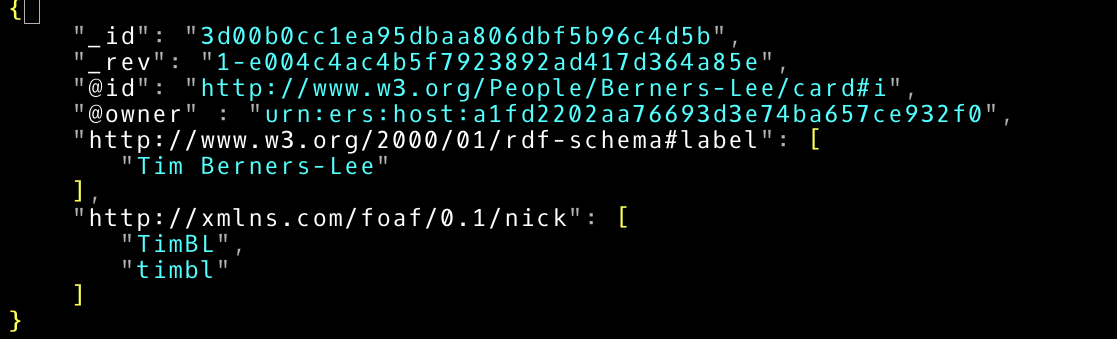}
\caption{Predicates as keywords \label{overflow}}
\label{fig:jsonpreds}
\end{figure}

\begin{figure}[ht!]
\centering
\includegraphics[width=60mm]{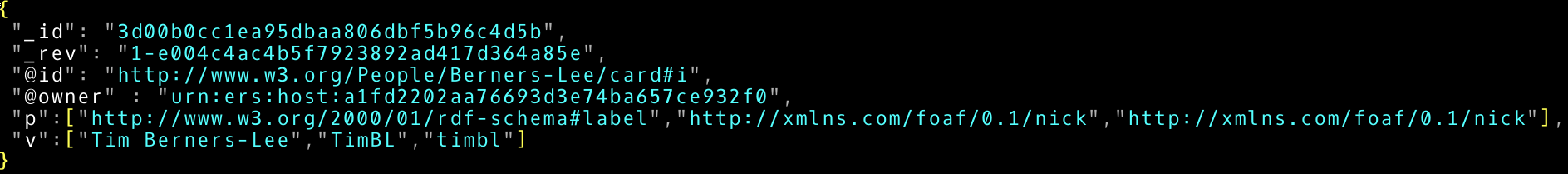}
\caption{Synchronized arrays \label{overflow}}
\label{fig:jsonarr}
\end{figure}

ERS uses CouchDB as the storage back-end and takes advantage of the replication mechanisms it provides in order to transfer data between peers. However, the data model of the database is JSON documents. In order to transform RDF tuples into JSON two options have been considered: predicates as keywords and synchronized arrays (Figure\ref{fig:jsonpreds} and Figure\ref{fig:jsonarr} respectively).

In \cite{ersdevs}, the disk space occupied by the two options in CouchDB is compared and it is concluded that this should not be a deciding factor, since no major differences were observed. The option of using predicates as keys was chosen because of its similarity to JSON and because it facilitates data manipulation.

The next issue is that of mapping RDF entities to documents. In \cite{ersdevs} a study of the various options has been performed and one document per graph entity combination was chosen as the best option. In this case each node that makes statements about an entity will create a local document containing those statements. Read performance will decrease (when compared to having one document per entity) because all the pieces that make up the same entity need to be merged by applications using ERS. However, this simplifies writing and synchronization as CouchDB natively supports the required operations.

\subsection{Data stores}

\begin{figure}[ht!]
\centering
\includegraphics[width=60mm]{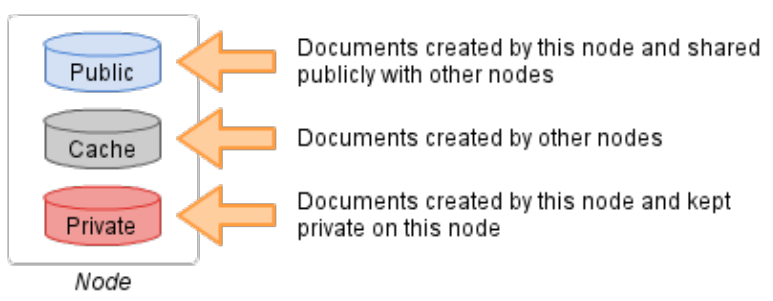}
\caption{The three data stores present on an ERS contributor. The private store will not be readable by others, as opposed to the public store which can be searched and whose updates will propagate over the network. The cache is used for storing previous user queries and in the replication mechanism. \cite{ersdevs} \label{overflow}}
\label{fig:datastores}
\end{figure}

Each ERS contributor has access to three data stores (represented internally by three different CouchDB databases) : public, private and cache (Figure\ref{fig:datastores}). The private stores represents the collection of documents that the user does not want to be visible by any other nodes in the system. Examples of data that is stored here is personal private information for participants to a conference or sensitive business information in the case of a vendor.
The public store holds statements that can be queried by others and that get replicated by bridges. Examples include a public list of skills and work experience that a conference participant wants to share with others or a list of products and their prices for a vendor.

The cache contains information that is not local to a node, but has been queried by him. In the case of a vendor, he can have in his cache documents related to prices of other vendors that he does business with, which get updated every time there is network connectivity between them.

\subsection{Synchronization}
\begin{figure}[ht!]
\centering
\includegraphics[width=60mm]{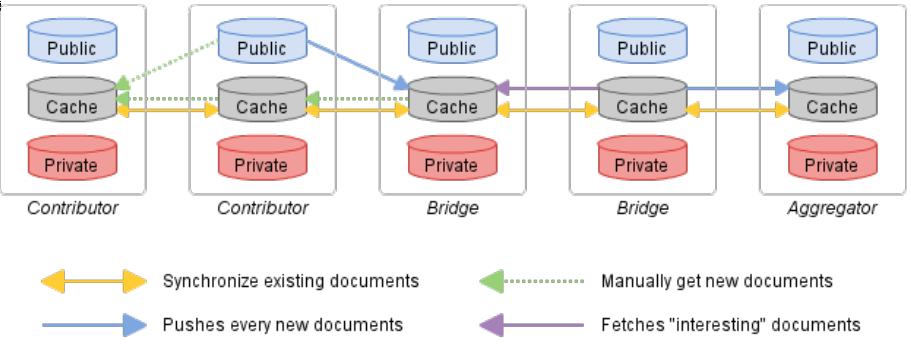}
\caption{Synchronization protocol. Updates for documents already in the caches propagate to all nodes who have those documents in their own caches. New document behavior varies by synchronization type.  \cite{ersdevs}\label{overflow}}
\label{fig:synchronization}
\end{figure}

The synchronization model proposed by ERS consists of the following parts: contributor-contributor, contributor-bridge, bridge-bridge and bridge-aggregator. In \cite{ersdevs} an overview of the synchronization schema used is given. A graphical description can be seen in Figure\ref{fig:synchronization}.

\begin{enumerate}
\item Contributor-Contributor synchronization. When two contributors are on the same network, but without a bridge, this synchronization protocol is used. In this case, their caches are synchronized through a filtered replication. Each node will only receive updates about the documents that it already has in his cache. This avoids the issue of a contributor with a very large cache filling up others' databases with data they are not interested in.
\item Contributor-Bridge synchronization
When contributors are connected to a bridge, they drop synchronization links to other contributors. This reduces the total number of connections from quadratic to linear (in the number of contributors). Bridges keep all the public documents of contributors in their cache (through an unfiltered continuous synchronization). Contributors receive from bridges updates for the items they have in their caches through a filtered replication.
\item Bridge-Bridge synchronization
Bridges exchange updates for the documents in their cache store and have the ability to get new documents that contain statements about entities already found in their cache.
\item Bridge-Aggregator synchronization
The Aggregator receives updates from the caches of all the bridges that it is connected to and it can provide a read-only global snapshot of the public data in the system.
\end{enumerate}

\subsection{Implementation overview}
This subsection focuses on ERS nodes and bridges, because the usage scenarios that are presented in this paper do not require the presence of an aggregator.

There are two main components of ERS nodes and bridges: the API and the daemon. The daemon deals with discovering peers, implementing replication, handling the various configuration options and logging. The API is designed to be interacted with by the users of ERS. It exposes methods for creating, deleting or retrieving entities, adding statements, querying remote nodes and searching.

The nodes and the bridge share the same implementation, with the main difference being the replication strategy that is used. The choice between which component to run on a node can be made in the ERS configuration file.

\textbf{ERS API.}The main purpose of the ERS API is to provide a convenient means of working with CouchDB in a linked data model, and operates on Documents and Entities.

	\textbf{Document.} An ERS document provides a wrapper around CouchDB documents. It is represented by a 	Python dictionary that will map directly to a JSON format when written to the database. The Document class has methods for adding, removing and modifying tuples (entity, predicate, value).

	\textbf{Entity.} The Entity class is a wrapper around the different documents that compose the description of an entity. These documents can be from either the local storage(public, cache and private) or from remote locations(such as bridges or other peers). The methods that this class exposes handle adding, modifying or deleting statements about the entity in a particular scope(public or private), as well as getting all the tuples that are associated with it.

The API exposes methods that Search and Retrieve particular entities either by their name or by looking for property-value matches. It also allows creating, updating and deleting entities, moving items to and from a node's cache and communicating with the daemon about replication updates.

\textbf{ERS Daemon}.The main purpose of the ERS daemon is to seamlessly abstract the network part of the system. The connection to other peers is established by using Avahi \footnote{\text{http://www.avahi.org/}}. Avahi is a LGPL implementation of the Multicast DNS and DNS Service Discovery specifications of the zeroconf network protocol \cite{zeroconf}. When the daemon is started, it reads the configuration file, and sets the peer type to be either a bridge or a contributor.  After setting up logging, it publishes the node as a network service on the local network. It is worth mentioning that because service names must be unique(in order for nodes to be able to see all other nodes), the host name of a computer gets a randomly generated string of characters appended to it.

The daemon also detects all the other nodes that are on the same local network as the host. It then tries to setup the replication policy, by creating appropriate documents in the CouchDB replicator database(linking local databases to those of peers or of the bridge if it is present). This replication can also be filtered to only include documents in a contributor's cache.

\section{ERS testing framework}
This section describes the changes that were made to ERS in order to make deployment and testing fully automatable. Automation and being able to programatically create complex scenarios helps validate the system before deploying it. To this end, the main objectives of this work can be summarized in the following research questions:
\begin{itemize}
\item{} Can the orchestration of an ERS deployment be automated, in order to create complex simulations of real-world deployments of ERS?

\item{} How does ERS perform in realistic scenarios that exhibit  less than ideal conditions ? What are the minimum hardware and network connectivity requirements ?

\item{} Does the ERS scale to a large number of users?
\end{itemize}

\subsection{Testing environment setup}

In order to automate the ERS deployments, some parts of it needed to be modified in order to allow an external application to issue commands to it.

The ERS API consists of a python module that can be imported into a client application and a command line interface that can be used directly from a machine that has ERS installed. An HTTP API has been developed that implements most of the functionality. This would simplify building a graphical user interface to simply building a web page that sends requests to the locally running web server. However, the main use of the HTTP API in this project was to enable sending of commands remotely. This was done in order to be able to construct automatic testing scenarios that could enable the orchestration of an entire deployment.

The daemon needs to expose a way to receive commands from the ERS API (in order to trigger replication updates). Multiple options were considered
\begin{itemize}
\item{}Merging the API and the daemon in order for them to call each other. This is undesirable because the API and the daemon are separate logical parts of the program
\item{}Signal communication between them. Does not work because receiving a signal causes dbus to exit its main loop (regardless of whether the signal is masked)
\item{} Including a lightweight HTTP server in the daemon that will accept commands.
\end{itemize}
Ultimately, the HTTP server option was chosen because it preserved the logical separation between the API and the daemon and was easy to use.
This HTTP server also exposes methods for stopping and starting the HTTP daemon, for use in automated testing.

\subsection{Configuration}
Some of the default settings of COUCHDB needed to be changed for the correct functioning of ERS. The option to delay writes to disk was disabled, as per the recommendations of the COUCHDB manual. The number of replication workers has also been increased so that if there is a node which has a very large amount of data to transfer, it does not smother all the others. Having multiple threads reduces the chances that all of them are busy transferring very large databases (for example from a node which has been offline for a very long time).

\subsection{Virtualization and simplifying deployment}
ERS was designed so that a single instance is running on a machine. In order to cover a larger scale when testing, doing it on physical hardware was difficult. First of all, ERS had to be manually configured and installed, including each dependency(CouchDB, avahi, dbus, etc.). In order to address this issue the Puppet automation tool can be used. A puppet manifest has been developed that automates the setup process of ERS and simplifies usage in cases where the other deployment options described below are unavailable (for example because the hardware is low-end and cannot run virtual machines or docker images are not available for the target platform). If the puppet agent is installed on the system, the manifest handles all the configuration and installation of ERS.

Virtual machines have been used to develop and test ERS. These permit running multiple instances of the system on the same machine in order to investigate communication behavior. These virtual machines can be assigned public IP addresses so that other nodes on the same network (including those running on different hardware) can interact with each other. In order to simplify the management of the various virtual machines, Vagrant has been used. A new virtual machine that can run ERS can be spun up in minutes, using puppet for provisioning. This has allowed basic interaction tests (such as a few nodes connected to a bridge).

Although virtual machines simplify the development, they contain many pieces that the ERS does not entirely need(such as a completely independent file system and network stack, etc.). Because of this it is also very resource-demanding to run multiple virtual machines on the same physical hardware. To address this issue, Docker\footnote{\text{https://www.docker.com/what-docker}} has been used. Docker provides a convenient means to use Linux containers and produce isolated deployments of ERS, while sharing operating system components. This has made it possible to scale the number of ERS instances running on a desktop computer from 5-6 to 30. The containers are also given public IP addresses so that they can communicate with other ERS instances, through the use of pipework\footnote{\text{https://github.com/jpetazzo/pipework}}.

A comparison between containers and virtual machines can be seen in \cite{containervsvm}. Because they are much simpler to set up and can offer similar if not better performance, they have been used as the deployment method for the more complex experiments that were performed on ERS. However, all these three methods of running ERS can inter-operate. A deployment consisting of a raspberry pi 2 model B running ERS natively, a laptop running 4 virtual machines with ERS and a desktop computer running multiple ERS containers has been tested using a local network, and all instances propagate statements as desired.

\subsection{Poor network conditions simulation}

In order to observe the behavior of ERS under less than ideal network connectivity, a basic implementation of the Netflix Simian Army suite of tools has been developed. This consists of programs that can cause different types of issues within a running distributed system. The original monkeys were designed to run on AWS and could not be run locally. Also, the initial setup work creates an entry barrier that was unnecessary for the scope of the experiments presented in this work. Since the orchestration of the ERS deployment mainly used virtual machines and docker containers, the tools that were created were designed to control instances created in this way.

The chaos monkey randomly terminates instances within the running pool of nodes. This has the purpose of simulating abruptly losing connectivity or power to one of the nodes or bridges.

The latency monkey injects packet loss, corruption, latency, duplication and reordering in a running node. Since the target of deployment for ERS is situations where perfect connectivity is unlikely, this should help create an impression of the behavior of ERS in such an environment.

This rudimentary implementation is available on github \footnote{\text{https://github.com/grameh/monkeybusiness}}, and is designed to control virtual machines and docker containers running locally. To interface with them, it uses the python-vagrant package \footnote{\text{https://pypi.python.org/pypi/python-vagrant}} for Vagrant and the docker-py package \footnote{\text{https://github.com/docker/docker-py}} for Docker. The chaos monkey simply uses these clients to turn nodes off and on. The latency monkey uses the clients to issue commands which take advantage of the advanced traffic control features available in the linux kernel \cite{linuxtraffic}. Although currently they have only been tested with locally running virtual machines and containers, the functionality could be extended by connecting the vagrant and docker modules to remote clients.

\section{Evaluation}
This section describes the various experiments that were performed during the development of ERS in order to verify functionality, scalability and fault tolerance. The first subsection summarizes the deployment and experimental setup. The following one describes a simple experiment with two nodes running on virtual machines on the same physical machine, with ideal network connectivity. The last two subsections describe two simulations of deployment scenarios: a social network platform for conferences and a deployment scenario for disconnected remote villages that still need to share data (in this case, price lists of different stores).
\subsection{Experimental setup}
To summarize section 4, almost all of the aspects of an ERS deployment can be interacted with remotely (e.g. from a central machine that orchestrates more complex test scenarios). The topology can be modified by starting new node or bridge containers through docker, and the network connectivity between these instances can also be adjusted through pipework. Within a particular node itself, the daemon can be turned on or stopped by issuing a HTTP request to the daemon web server. The API can also be controlled through HTTP requests (e.g. in order to add, remove, search for entities or statements). Network quality can be controlled through the use of the latency monkey, and arbitrary failure can be simulated with the help of the chaos monkey.

\subsection{Functionality tests}

To facilitate deployment and avoid breaking functionality while modifying various pieces of ERS, a suite of basic tests has been developed. These target the creation of new entities, modification of existing ones as well as deletion. Search and retrieval of entities also constitute "core" functionalities and thus are included into the suite.

This set of tests has been included into the development process by running them on every commit with the help of the Travis continuous integration service.

Rudimentary performance testing has been performed on the command line interface. Because it persists on every operation, it introduces a certain overhead when interacting with CouchDB. The number of new documents that can be created on a machine with low-end specifications is approximately 3/second. The number of new statements about an entity that can be made each second is approximately 10. However, this is considered to be a reasonable amount for the use case. If better performance is required, the ERS API module can be included in the client application, which provides more granular methods for writing to the database (for example building an entity in-memory and committing all the documents in a single API call). Table\ref{table:1} shows the performance of ERS under a virtual machine with 1 core and 512MB of RAM.

\begin{table}
\begin{center}
 \begin{tabular}{||c c ||}
 \hline
 Type & Operations/second \\ [0.5ex]
 \hline\hline
 entity creation & 5  \\
 \hline
 property edits & 20  \\
 \hline
 value edits & 20 \\ [1ex]
 \hline
\end{tabular}
\caption{CLI Operation throughput on a raspberry pi 2 model B}
\label{table:1}
\end{center}
\end{table}

\subsection{Simple experiment}
In this experiment the basic replication behavior will be observed. Two virtual machines have been connected through a private network. Both machines create documents that describe the same entity. One machine looks up if others have statements about it, and decide to cache the entity. This triggers the daemon's replication protocol. Updates are written continuously on the second machine and the propagation behavior is shown in Figure\ref{fig:replication}. Node1 (whose cache is being read) is represented by the blue line and node2 (which makes statements) is represented by the orange line.

\begin{figure}[ht!]
\label{fig:replication}
\centering
\includegraphics[width=60mm]{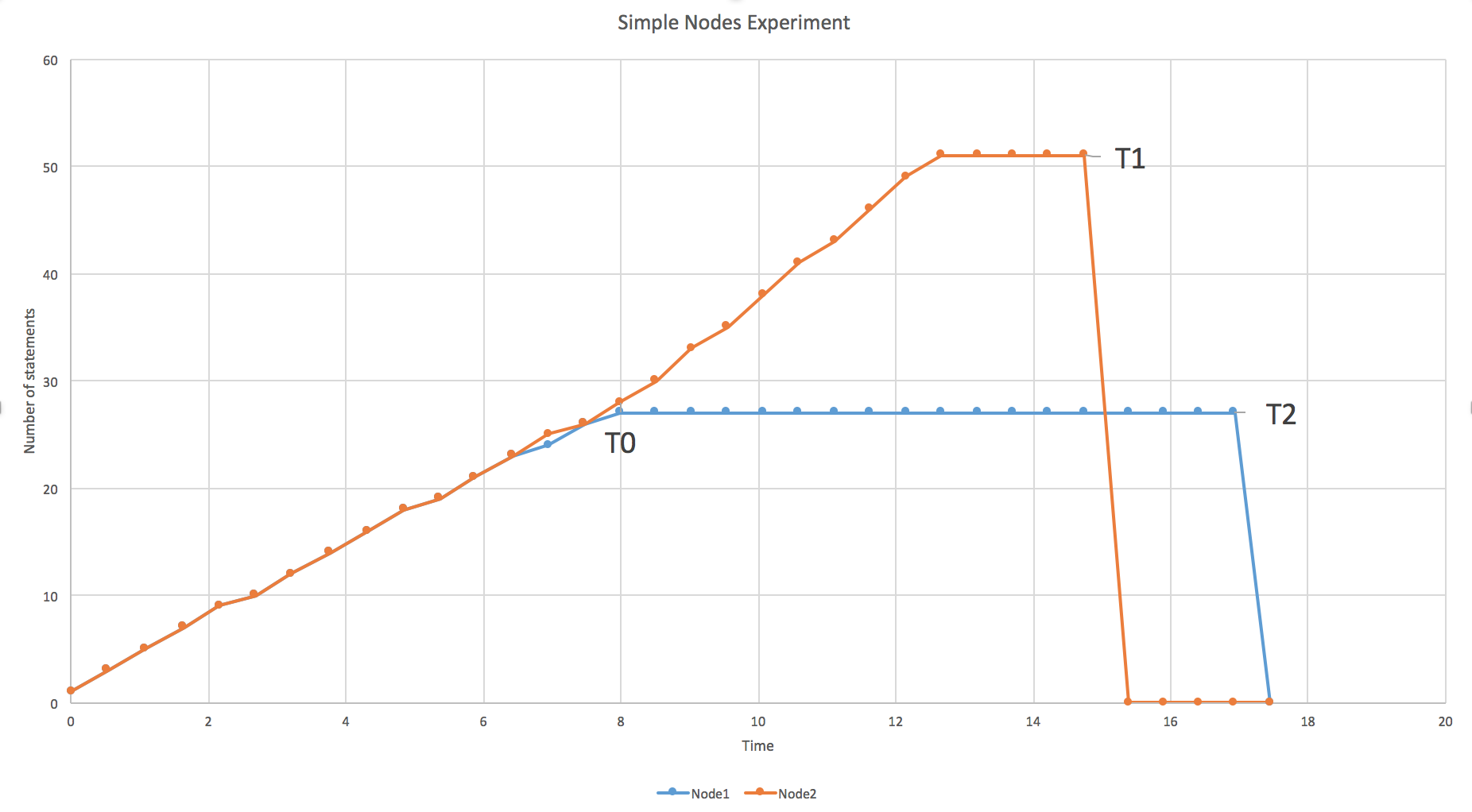}
\caption{Replication experiment. Node1 (whose cache is being read) is represented by the blue line and Node2 (which makes statements) is represented by the orange line. At T0, Node1 loses connectivity. Node 2 continues writing and decides to delete its document at T1. Once Node1 comes back online (at T3), the updates get propagated.\label{overflow}}
\label{fig:replication}
\end{figure}

In this particular experiment,at the time point T0, node1 loses connectivity. Updates to node 2 are still written but they cannot propagate to node1. At T1, node 2 decides to delete the document containing the statements. Since node1 is still offline, he does not receive this update.

At the time point T2, node1 is reconnected and almost immediately CouchDB propagates the new changes, thus removing the document from node1's cache. Note that node1's public statements do not get deleted from his machine, but only node2's statements that node1 had in his cache.

This experiment has been repeated with a bridge in between the nodes to check if it influenced replication, and the behavior was not modified. These simple experiments have been used throughout development as smoke tests for basic verification of functionality, especially after tweaks in the replication strategy.

\subsection{Conference}
The purpose of this experiment is to simulate how ERS would perform as a substitute for a social network (such as LinkedIn) in the context of a conference. In this scenario, there are various conference attendees with public profiles, who list their skills and workplace. Attendees can endorse each other for their skills. This scenario contains a static bridge node and various contributor contributor nodes that randomly connect and disconnect, with varying network quality.

Initially, each attendee has on his laptop his public profile. Once he connects to the local network of a particular conference room, he can initiate a search for other participants. He can endorse any other participant for his skills, and these endorsements will propagate to the bridge and thus are visible to all the other attendees.

\begin{figure}[ht!]
\centering
\includegraphics[width=80mm]{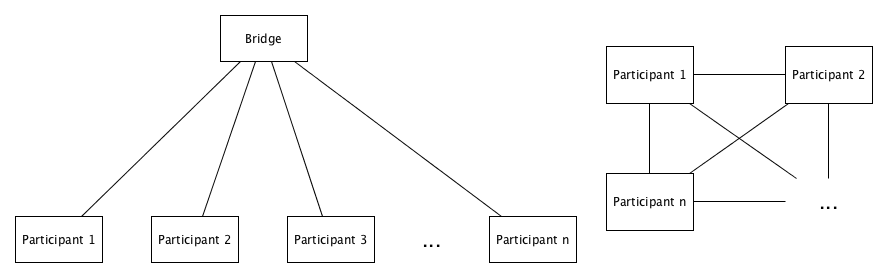}
\caption{Connectivity in case bridge is present (left side) and when it is not (right side). If there is no bridge present on the network, each ERS node connects to all the others on the same network, resulting in a quadratic number of connections. The bridge reduces the number of required connections from quadratic to linear.\label{overflow}}
\end{figure}

The experiment begins by starting the docker containers. Each container is based on the Ubuntu 14.04 LTS operating system, though any major Linux distribution can be used instead. Each container runs the software stack(dbus, avahi, CouchDB, ers daemon and ERS web API) and is given a public IP. It is important to note that because of the way peers are discovered(avahi discovers each of them as a separate "service", one at a time) there is a delay before they are all visible to each node. This delay increases linearly in the number of peers or bridges if they are present, from a negligible amount in case of 1 peer/bridge to approximately 3 seconds in the case of 40 peers/bridge. Each container then executes a search using the API for entities that are tagged with the property "ers:ConferenceAttendee", and caches the results. This allows CouchDB's automatic synchronization protocol to pick up statements made about those entities automatically. Each contributor then proceeds to make a random number of statements about others.

Monitoring the status of all the peers is done in a separate thread that queries each node for the percentage of documents found in its cache against the total number of documents created in the systems.

Through this experiment, we observe that the system scales to a larger number of users. With a bridge on a Raspberry Pi 2 model B, it was able to correctly propagate statements between contributors with as much as 40 concurrent contributors.

\subsection{ERS Bridge on a truck}
\begin{figure}[ht!]
\centering
\includegraphics[width=60mm]{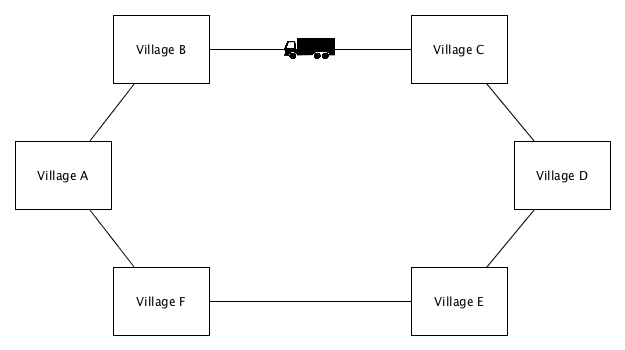}
\caption{Mobile bridge experiment. In this experiment, each village has an ERS node, and the truck has an ERS bridge. The truck passes through the villages in clockwise order, propagating updates among them. \label{overflow}}
\label{fig:truckexperiment}
\end{figure}

This experiment aims to simulate an ERS deployment in a remote location that has no access to the internet, thus removing the assumption that any part of the system (node or bridge) is static and always connected. In Figure\ref{fig:truckexperiment} 6 remote villages that each contain a store, and a truck that periodically passes through all of them can be seen. The experiment assumes that in each village there is a vendor that would like to sell or buy products from neighboring villages, and their only means of daily communication and updates is the truck driver that passes through.

In this scenario, we would deploy in each village a contributor node of ERS and on the truck, a bridge node. Each vendor controls the list of items that he is selling and the prices associated with those items in an ERS public document. These documents get uploaded to the bridge the first time it comes in contact with them. Vendors can search for the nodes they are interested in and choose whether they want to cache those entities.

If vendors make modifications to their prices or list of items, these modifications get picked up when the truck passes through his village. Also, when the truck arrives, if he has other vendors in his cache, he will automatically get updated versions of their offerings if the truck stays for at least 5 seconds (this can be as low as 1 second if network connectivity is very good).

Having fixed the number of seconds that the bridge is connected to peers to 5 seconds, investigations have been made into the behavior of ERS in less than ideal conditions. Firstly, the Latency Monkey injects artificial latency into each packet that leaves the system. The system can perfectly handle 100ms of delay(each way) and within the allocated time the bridge and the node will completely synchronize. If latency goes over 125ms each way, the replication procedure will not be able to finish in time for all the nodes.

The next characteristics of poor network connectivity that were investigated are packet loss and packet corruption. In this case the system can tolerate up to 15 percent of the packets being lost or corrupted on each end (the contributor and the bridge). A value that is higher than that will cause rapid degradation in the percentage of statements that are synchronized (in the given 5 second window).

Duplicating packets does not seem to have an important effect on the performance of ERS, as no difference has been observed with as much as 60 percent duplication.

There are a few more issues that are considered and that ERS can handle. If a, perhaps competing, vendor wants to modify the price list of a different node, he will instead create local documents with the modifications. These do not get merged automatically with the original ones because the CouchDB documents will have different ids. Thus, if node A has only the document of node B in cache, the automatic replication will not get the changes that anyone other than node B makes.

In this experiment, the topology was constantly shifting, in the sense that the bridge only had connectivity to contributors for brief amounts of time. We have observed that even with poor connectivity, the system was robust, and propagated the updates.

\section{Discussion}

The results of the experiments allows the answering of the research questions.

\subsection{Can the orchestration of an ERS deployment be automated, in order to create complex simulations of real-world deployments of ERS? }

Having made all of the components of ERS remote controllable and the deployment automatable, we were able to create simulations of more complex scenarios. This can allow the creation of smoke testing environments that verify ERS behavior, and run as part of a continuous integration system. The benefits of doing this are two-fold. Firstly, having test scenarios more complex than unit tests and modeled after realistic use cases provides some guarantee of desired functionality. Additionally, having these tests repeatable and running them continuously gives developers confidence to make changes without worrying whether they will break the experience users. It could be argued that in a system designed to function in remote areas, where updating is non-trivial if not impossible, this is extremely important.

\subsection{ How does ERS perform in realistic scenarios that exhibit  less than ideal conditions ? What are the minimum hardware and network connectivity requirements ?}

The two real-world scenarios simulate realistic deployments of the platform, with different objectives. The conference test verifies that a stationary bridge can handle contributors that connect, disconnect and interact with each other in a continuous fashion. The village test demonstrates that bridges need not be stationary and that various completely disconnected parts of the system can communicate through the bridge in a reliable and secure fashion. These two behavior investigations suggest that ERS is ready for field-testing.

Within the conference experiment without artificial network degradation, each node reports 100 percent completion during each query. This has been tested for conference sizes up to 30, as this is the maximum number of ERS containers that can be launched on a desktop system with 8GB of RAM.
This suggests that the ers synchronization protocol can easily handle a considerable number of concurrent peers in this use case.

The villages experiment, in which the bridge is mobile and synchronizes data across multiple disconnected areas, showed that under perfect connectivity (docker containers within the same machine) replication happens very fast. As long as the bridge was on the same network as a contributor for at least half a second, the small data set would be transferred without any noticeable delay. More interestingly, the overhead of using the command line interface or the web API means that CouchDB can replicate faster than nodes can write to ERS.

The correctness of ERS functionalities has also been verified and a suite of tests has been created. Storage and API behavior is verified through unit tests. Communication and correctness of the functionality of the ERS daemon can be verified through automated testing that use virtual machines.

ERS can be deployed and used on low-end hardware. Both ARM and x86/x64 CPU architectures are supported, and even low-end hardware such as the Raspberry Pi can easily handle running the system.
Having investigated the effects of the hallmarks of poor network connectivity (latency, loss, corruption, duplication) on the functioning of ERS, we can conclude that the system is resilient with respect to low quality connections. The system also handles being offline gracefully, with the only downside being that the user sees stale data.

\subsection{Does the ERS scale to a large number of users? }

The original ERS project had been tested with at most 4 XO physical laptops connected with a bridge on a raspberry pi, with more or less ideal network connectivity (local Ethernet connection). This work has presented experiments with a combination of simulated devices and physical hardware, with up to 40 concurrently running nodes and with artificially bad network connectivity. The scenarios which ERS targets, such as students in a classroom, remote area conferences, stores with no internet access, should see no performance issues.

\section{Conclusion}
The goal of this paper was to investigate whether it was possible to setup ERS for automated testing, in order to decide if was ready for usage. Through the various functionality tests the correct functionality of various components can be assured. The tests on virtual machines verify that communication between nodes works as expected. The conference test indicates that it can scale to a larger number of uses and the truck test shows the resilience with respect to poor network connectivity.

Future works should expand the scope of the scalability testing by investigating, for example, how many contributors can be connected to a bridge without losing functionality, or how peers from different geographical locations connected through the internet influence the behavior of the system. An implementation of the aggregator should be finalized and integrated with the other components.

It is usually the case that tests in controlled environments differ from real-world scenarios. Investigating the behavior of the system in an actual deployment should be carried out.

\end{document}